\documentclass[preprint,nofootinbib]{revtex4-1}
\usepackage{amsmath}
\usepackage{verbatim}
\usepackage{mathrsfs}
\usepackage{amsfonts}
\usepackage{graphicx}

\begin{document}

\author{Pedro Aguilar}
\email{pedro.aguilar@nucleares.unam.mx}
\affiliation{Instituto de Ciencias Nucleares, Universidad Nacional Aut\'onoma de M\'exico\\ Apartado Postal 70-543, Ciudad de M\'exico 04510, D.F., M\'exico}

\author{Yuri Bonder}
\email{ybonder@indiana.edu}
\affiliation{Physics Department, Indiana University\\
727 E. Third St., Bloomington, Indiana, 47405, USA}
\affiliation{Instituto de Ciencias Nucleares, Universidad Nacional Aut\'onoma de M\'exico\\ Apartado Postal 70-543, Ciudad de M\'exico 04510, D.F., M\'exico}

\author{Daniel Sudarsky}
\email{sudarsky@nucleares.unam.mx}
\affiliation{Instituto de Ciencias Nucleares, Universidad Nacional Aut\'onoma de M\'exico\\ Apartado Postal 70-543, Ciudad de M\'exico 04510, D.F., M\'exico}

\date{IUHET 571, March 2013; Published in Phys. Rev. D 87, 064007, 2013}

\title{Experimental search for a Lorentz invariant spacetime granularity: Possibilities and bounds}

\begin{abstract}

\noindent We consider a search for phenomenological signatures from an hypothetical space-time granularity that respects Lorentz invariance. The model is based on the idea that the metric description of Einstein's gravity corresponds to a \emph{hydrodynamic} characterization of some deeper underlying structure, and that Einstein's gravity is thus to be seen as emergent. We present the specific phenomenological model in detail and analyze the bounds on its free parameters established by an experiment specifically designed to test this model.

\end{abstract}

\maketitle

\section{Introduction}\label{intro}

\noindent The notion that space-time at the fundamental level might be endowed with some granular structure has been considered since the first attempts to combine general relativity and quantum theory. For most of the last century there existed among physicists a widespread conviction that, even if such granular structure were real, there was no hope of ever observing its effects. This is essentially because the natural scale for such granularity is the Planck length $l_P=1.6\times 10^{-35} \ \rm{m}$ which is many orders of magnitude smaller than the scale that can be directly explored today or in the foreseeable future.

This pessimistic outlook was transformed by the observation that such space-time granularity could manifest itself through a conflict with exact Lorentz invariance. There had been, in fact, an ongoing research project devoted to precision tests of fundamental symmetries, such as Lorentz Invariance and CPT \cite{Kosteleckyinitial},  but  an  important surge in interest came to the field due to the observation that what seemed as the natural estimates of the  strength of the Lorentz Violating effects resulting from a Planck scale granularity were within the experimentally accessible region, in the case of gamma  ray bursts with origin at cosmic distances\cite{A Camellia}. The basic idea is that, at the phenomenological level, the space-time granularity would become manifest by a modification of the dispersion relations of massive and massless particles, which, for example, could take the form $E^2 =p^2 + m^2 + \xi E^3 / M_P$, where $\xi$ is a free phenomenological parameter and $M_P$ is Planck's mass. Soon 
after this, other possibilities for testing this idea were proposed \cite{Kozameh-Gleiser, Jacobson, Mattingly} and led to more stringent bounds on parameters such as $\xi$. At the same time, colleagues working on various approaches to quantum gravity produced schemes were such effects would emerge ``naturally" (see \cite{Gambini, Alfaro1999, Morales2002, Alfaro2005, Alfaro2005-2, Urrutia:2006} for proposals based on Loop Quantum Gravity and \cite{Kosteleckyinitial, KosteleckyPotting, Ellis2000, Ellis2002} for proposals 
inspired on String Theory).

At that point, the possibility of obtaining bounds on parameters like $\xi$ was either connected with very high energy particles or with very high precision measurements (such as the Hughes-Drever experimental tests of isotropy of physics in the laboratory \cite{HD, HD Bounds}). Some of us were led to consider a realm where these two aspects come together: radiative corrections in quantum field theory. These corrections involve the dynamics of virtual particles of all possible energies and its effects can be measured with exceedingly large precision (like the famous anomalous magnetic moment of the electron). The general view is that the effects of those high energy loop corrections can be absorbed into the renormalization of the coupling constants, but as the example of the anomalous magnetic moment of the electron shows, sometimes there are interesting nontrivial remnants of renormalization. These remnants are particularly important when symmetry issues are involved. An investigation on these issues showed 
that the effects of a space-time granularity that is tied to a preferential frame and that induces a cutoff in the field's momenta would be so large that they would have been observed long ago \cite{Collins2004}. These findings have been disputed and arguments have been put forward specially by people working in Loop Quantum Gravity\footnote{In Loop Quantum Gravity, these results might severely constrain the kind of states that can be considered as suitable characterizations of space-time, even Minkowski space-time \cite{Gambini-Pullin}. On the other hand, Loop Quantum Gravity by itself does not require any breakdown of Lorentz invariance \cite{Rovelli}. Nevertheless, it is fair to say that, to our knowledge, researchers in that field have not been able to construct a state where these effects are clearly absent.}. These arguments have in turn been replied \cite{Polshinski2011} and, as things stand today, we do not believe there is any argument effectively contesting against the generic considerations 
indicating that, in absence of an unbroken custodial 
symmetry, 
all possible Lagrangian terms (consistent with the unbroken symmetries of the theory) if not introduced from the beginning, are generated through radiative corrections. In our case, those would involve the kind of unsuppressed Lorentz violating terms uncovered in Ref. \cite{Collins2004} which indicate that different particle species would have different light cones, something that is excluded with a very wide margin by observations \cite{Coleman}.

This argument motivated us to take a more agnostic approach to the question of how a discrete space-time structure may become manifest. In doing so, we avoid committing to any particular proposal for a quantum theory of gravitation, but we use some of the experience gained in the course of the above mentioned searches. The two main lessons we incorporate in our constructions are that the naive analysis that gave order of magnitude estimates can be underestimating the level at which the granular structures manifest themselves (as we have seen, the granular structures associated with the breaking of Lorentz invariance led to effects which were NOT suppressed by the Planck scale), and that the possible effects should be characterized at the level of an effective theory, in terms that might not be the most natural ones (as in the observation that the change in the dispersion relation was not an appropriate characterization of the emergence of specific effective metrics, as found in Ref. \cite{Collins2004}).

The basic idea is that our current geometric characterization of space-time geometry might be nothing more than an effective description emerging from an underlying structure similar to the hydrodynamic characterization of a liquid by the Navier-Stokes (N-S) equations. According to this view, gravity as is currently described by general relativity, would be just an emergent phenomenon while the underlying substrate represents the gravitational degrees of freedom at the quantum level. In other words, we consider that general relativity should be regarded as some sort of \textit{hydrodynamical} limit of a more fundamental theory in which the description of space-time structure does not rely on a metric tensor or closely connected variables. Hydrodynamics is the effective theory describing the long wavelength, large frequency, low energy modes of highly coupled systems with a large number of ``particles", \textit{i.e.}, fluids. We know that such description is appropriate for describing the fluid at length 
scales larger than the mean free path of the particles that make up the fluid, and time scales larger than their mean free time. At those scales, only a small number of degrees of freedom are necessary to characterize the state of the system: the \textit{hydrodynamic} ones. In the case of a perfect fluid, for example, it can be described by the energy density, the entropy density and the field of velocities; all related by the Euler, the continuity and the thermodynamic equations. (Note that it was not necessary to introduce the concept of a fundamental constituent --the particle-- in order to describe the fluid, of which the XVIII century physicists who developed the theory were unaware.) The point of view we taken here is that something similar occurs with the gravitational interaction where the characteristic physical scale could be determined by $M_P$. That is, in fact, the situation in various approaches to quantum gravity such as Loop Quantum Gravity or the Poset proposal \cite{Posets-Sorkin}. This 
point of view seems 
to 
be further supported by the analogy between the behavior of black holes and the laws of thermodynamics \cite{Bekenstein:1973ur}, and by the ideas developed in Ref. \cite{Jacobson:1995ab} (see also Refs. \cite{Volovik2003, Padmanabhan:2003gd, Hu:2005ub}).

Now let us characterize the general nature of our approach, even if only schematically, within the general current understanding of physical theory. As it is well known, when following canonical quantization procedures such as the old Wheeler-DeWitt proposal \cite{WdW}, or its more modern incarnation in the form of Loop Quantum Gravity \cite{RovelliLQG}, one ends up with an atemporal theory. This is known as the problem of time in quantum gravity \cite{WdW,Isham:1992ms}. That is, in both schemes one starts with a formulation in which the basic canonical variables describe the geometry of a 3-spatial hypersurface $\Sigma$, and characterize the embedding of this 3-surface in a 4-dimensional space-time. From those quantities one is led to the identification of a set of canonical variables, which we denote here generically by $({\cal G}, \Pi)$.
The problem is that time, or its general relativistic counterpart, a time function usually specified by the lapse function and the shift vector, simply disappears from the theory because the Hamiltonian vanishes when acting on physical states (those satisfying the diffeomorphism and Hamiltonian constraints). The issue is then how to recover a space-time description of our world, clearly an essential element if one seeks to connect the theory with observations. One of the most favored approaches towards addressing this problem is to consider, together with the geometry, some matter fields, which we describe here schematically by a collection of ordered pairs of canonically conjugate variables,
\begin{equation}
\lbrace(\varphi_1,\pi _1),\ldots, (\varphi_n, \pi_n)\rbrace ,
\end{equation}
and to identify an appropriate variable (or combination of variables) in the joint matter gravity theory that could act as a physical clock $ T(\varphi_i,\pi _i, {\cal G}, \Pi)$. The next step consists in characterizing the state for the remaining variables in terms of the correlations of their values with those of the physical clock. That is, one starts with the wavefunction for the configuration variables of the theory $ \Phi(\varphi_1,\ldots,\varphi_n, {\cal G})$, which must satisfy the Hamiltonian and momentum constraints $ H_\mu \Phi(\varphi_1,\ldots\varphi_n, {\cal G}) =0$. Next, one needs to obtain an effective wavefunction $\Psi$ for the remaining variables by projecting $\Phi$ into the subspace where the operator $T(\varphi_i,\pi _i, {\cal G},\Pi)$ takes a certain range of values. Let $P_{T, [t , t+\delta t]} $ be the projector operator onto the subspace corresponding to the region between $t $ and $ t+\delta t$ of the spectrum of the operator $T$, then, one attempts to recover a Schr\"odinger-like 
evolution equation by studying the dependence of $ \Psi (t) \equiv P_{T, [t , t+\delta t]} \Phi $ on the parameter $t$. After obtaining, by the above procedure, a wavefunction associated with the spectrum of $T$, it could be used to compute the expectation values of the $3$-dimensional geometrical operators (say the triad and connection variables of Ashtekar, or some appropriate smoothing thereof) for the wavefunction $\Psi(t)$. Such collection of quantities could be seen as providing the geometrical descriptions of the ``average" space-time in terms of the $1+3$ decomposition. In other words, one would have constructed a space-time where the slicing would correspond to the hypersurfaces on which the geometrical quantities are given by the expectations of the projected wavefunctions $\Psi (t)$, and thus, one would be able to characterize the space-time and its slicing in terms of the lapse and shift functions.

The precise realization of this procedure depends strongly on the situation and specific theory for the matter fields one considers. Such study is quite beyond the scope of the present paper, among other reasons because we do not have a fully satisfactory and workable theory of quantum gravity. On the other hand, several works along these lines already exist in the literature \cite{Pullin:2004}. The point we want to make here, however, is that in such setting the ordinary pseudo-Riemannian metric characterization of space-time is just an effective description, which emerges upon appropriate coarse graining and other suitable approximations from the underlying structure. There exists a large body of literature considering more generically the idea that space-time might be an emerging phenomenon \cite{Emerging}, and it is worth pointing that this seems to be required even in relatively conservative and canonical approaches.

The issue is that, once we accept this possibility, the idea that all effects of the underlying quantum gravity substrate might be characterizable in a simple form when expressing them in terms of the standard pseudo-Riemannian metric $g_{ab}$ would be far too simplistic. Of course we will only know this for sure once we have a well developed theory of quantum gravity, so in its absence we must avail ourselves to other examples which help us gain insights. This is where the hydrodynamical analogy comes in handy. There we have a case where, on the one hand, we have effective characterization via the N-S equation and, on the other hand, we have a fundamental theory of the constituents of the liquid which can, in principle, not only explain its simple properties encoded in the N-S equation, but can indicate when certain aspects might lead to observable effects and yet be indescribable in terms of the simple N-S equation and the standard hydrodynamical theory.

In this paper we present a phenomenological model of quantum gravity based on the point of view discussed above. One of the goals of this paper is to give a unified presentation of the model, in its final and consistent version, which is lacking in the literature. This is done in section \ref{model}. In section \ref{phen} we analyze the phenomenological consequences of the model and in section \ref{exp} we describe an experiment performed in order to test this model and we interpret the bounds obtained in such experiment in more detail. Section \ref{counterintuitive} is devoted to the discussion of some aspects of the model that might seem counterintuitive unless they are interpreted in the framework of effective phenomena emerging from a fundamental theory. Finally, we present some conclusions in section \ref{discussion}. The conventions we use include a $(-,+,+,+)$ signature for the space-time metric, Wald's \cite{Wald} convention for geometric tensors and the use of natural units.

\section{The model}\label{model}

\noindent As is discussed in the last section, we assume that if space-time has a granular structure at the fundamental level, it does not break Lorentz invariant. Not only do we not discard it but we hold Lorentz symmetry at a fundamental level as a guiding principle in our model. By taking this hypothesis seriously, a new kind of phenomenology of quantum gravity was proposed \cite{Corichi}. The idea is based on a simple everyday experience. Consider, for example, a building made of cubic bricks. Clearly, in any part of the building that does not share the cubic symmetry of the bricks, let's say, a circular wall, the presence of the bricks may be detected because it is not possible to construct a perfectly circular wall with cubic bricks. On the other hand, the symmetry of flat walls is compatible with that of the bricks and one cannot tell that the walls are made with cubic bricks by simply looking at them. Translating this idea to space-time, we can propose that the presence of a Lorentz invariant 
granular space-time structure may become manifest where space-time is not Lorentz invariant, namely, in curved space-time regions. Moreover, where space-time is flat the presence of a Lorentz invariant space-time granularity would not become manifest. In other words, in curved space-time regions some non-trivial arrangement of the fundamental space-time constituents must take place revealing its presence through non-standard interactions of matter and space-time curvature. We consider that it is important to explore and develop this heuristic approach into a full phenomenology model, so that experimental consequences be extracted through which Nature informs us of the right access to quantum gravity.

In this work, we focus in fermionic fields without self-interaction, thus, the interaction is assumed to be described at the effective level by Lagrangian densities of the form
\begin{equation} \label{Lagrangian}
{\cal L}= \sum_m e {\Xi}^m_{ab\ldots} {\bar\psi}_m O^{ab\ldots}\psi_m,
\end{equation}
where $e=\sqrt{-\det{g}}$, $m$ is a fermion flavor index, $\psi_m$ stands for the fermionic fields, $O^{ab\ldots}$ are some suitable operators constructed out of derivatives and the Dirac matrices (contracted with the tetrads $e_\mu^a$) and ${\Xi}^m_{ab\ldots}$ are tensors constructed locally and covariantly from the curvature and the free parameters of the model. It is easy to check that if $O^{ab\ldots}$ has $n$ derivatives, then the (mass) dimension of ${\Xi}^m_{ab\ldots}$ is $1-n$. For simplicity we limit our analysis to one fermion flavor and omit the label $m$.

To restrict the form of the interaction (\ref{Lagrangian}) we note that, according to Einstein's equations, the Ricci tensor at a space-time point $x$ is determined by the matter fields at $x$. Therefore, coupling the matter fields with the Ricci tensor can be considered, at a phenomenological level, as an irrelevant self-coupling. Thus, we focus on couplings of matter with the traceless part of the Riemann tensor, namely, the Weyl tensor $W_{abcd}$. Furthermore, the coupling terms must have dimensions such that they are not suppressed by a large power of a large constant with units of mass, which is taken to be proportional to $M_P$. As is pointed out in Ref. \cite{Corichi}, the obvious coupling term of dimension $5$ vanishes, outlining two possibilities, either studying phenomenologically uninteresting highly suppressed terms or constructing ${\Xi}_{ab\ldots}$ out of the Weyl tensor in a non-standard fashion. We choose to follow the latter approach.

The model initially described in Ref. \cite{Corichi} presented some problematic aspects which were tackled in following papers \cite{Yuri,Unambigous}. We here present the model in its final version which is well defined and free of ambiguities. The starting point of this construction is acknowledging that the space of space-time two-forms, $\mathcal{S}$, has a metric inherited from the space-time metric \cite{HallSCSGR},
\begin{equation}
G_{ab cd}=\frac{1}{2}\left(g_{ac}g_{db}-g_{ad}g_{cb}\right),
\end{equation}
that is, $G^{ab cd}$ is a map from pairs of two-forms into real numbers satisfying the properties of a metric. Note that the Weyl tensor ${W_{ab}}^{cd}$, being antisymmetric in its two pairs of indices, can be considered as a map from the space of two-forms into itself. We construct two self-adjoint maps out of the Weyl tensor
\begin{eqnarray}\label{3}
{{(W_+)}_{ab}}^{cd} &= &\frac{1}{2}\left({W_{ab}}^{cd}+{W^\dagger_{ab}}^{cd}\right),\\
{{(W_-)}_{ab}}^{cd} &= &\frac{1}{4}{\epsilon_{ab}}^{ef}\left({W_{ef}}^{cd}-{W^\dagger_{ef}}^{cd}\right),
\end{eqnarray}
where ${W^\dagger_{ab}}^{cd}$ represents the adjoint of ${W_{ab}}^{cd}$ and $\epsilon_{abcd}$ is the natural space-time volume $4$-form. The tensor field $\Xi_{ab\ldots}$ can be constructed out of the eigenvectors and eigenvalues of the self-adjoint maps and such is our proposal. We use $\lambda^{(\pm,l)}$ and $ X_{ab}^{(\pm,l)}\in \mathcal{S}$ such that
\begin{eqnarray}
\label{eigen map} {{(W_\pm)}_{ab}}^{cd} X_{cd}^{(\pm,l)}&=&\lambda^{(\pm,l)} X_{ab}^{(\pm,l)},\\
\label{planes} \epsilon^{abcd}X_{ab}^{(\pm,l)}X_{cd}^{(\pm,l)}&=&0,\\
\label{norm} G^{abcd}X_{ab}^{(\pm,l)}X_{cd}^{(\pm,l)}&=&\pm1.
\end{eqnarray}
The first equation is an eigenvalue equation where $\lambda^{(\pm,l)}$ and $X_{ab}^{(\pm,l)}$ are the eigenvalues and eigenvectors (or eigenforms) of ${{(W_\pm)}_{ab}}^{cd}$. The $\pm$ indicates to which self-adjoint map $\lambda^{(\pm,l)}$ and $X_{ab}^{(\pm,l)}$ are associated and $l$ labels the different eigenvalues and eigenforms. 

Observe that the Weyl tensor has in principle $6$ different eigenvalues and eigenforms corresponding to the number of antisymmetric pairs of space-time indices in $4$ dimensions. This also holds for ${{(W_\pm)}_{ab}}^{cd}$, suggesting that $l$ runs from $1$ to $6$. However, the identity
\begin{equation}
{\epsilon_{ab}}^{cd} {W_{cd}}^{ef} ={W_{ab}}^{cd} {\epsilon_{cd}}^{ef},
\end{equation}
which follows from Weyl symmetries, implies that if $Y_{ab}^{(\pm,l)}$ is an eigenform of ${W_{ab}}^{cd}$, so is ${\epsilon_{ab}}^{cd}Y_{cd}^{(\pm,l)}$. A similar relation holds for the Weyl adjoint, which in turn can be used to show that the maps (\ref{eigen map}) are always degenerated. The additional conditions (\ref{planes}) and (\ref{norm}) are conceived in order to discriminate from all these linear combinations; the chosen eigenforms are denoted by $X_{ab}^{(\pm,l)}$ and generically\footnote{The null eigenforms (with respect to $G_{abcd}$) are not well defined because equation (\ref{norm}) cannot be used to fix all the coefficients in the linear combination of degenerated eigenforms, therefore, these eigenforms must be discarded by assumption.} there are $3$ eigenforms, thus, $l$ runs from $1$ to $3$. Note also that the conditions (\ref{eigen map})-(\ref{norm}) are invariant under the replacement of $X_{ab}^{(\pm,l)}$ by $-X_{ab}^{(\pm,l)}$, which is problematic because an eigenform and its negative 
correspond to very different physical situations. Thus, $\Xi_ {ab\ldots}$ has to be (at least) quadratic in $X_{ab}^{(\pm,l)}$. Moreover, when space-time curvature vanishes, $\lambda^{(\pm,l)}=0$, and ${\Xi}_{ab\ldots}$ should also vanish in order to keep the model consistent with its motivation. 

Taking all the remarks discussed above under consideration and defining $\widetilde{X}_{{ab}}^{(\pm,l)}= {\epsilon_{ab}}^{cd}X_{cd}^{(\pm,l)}$ the proposed coupling term has the form
\begin{equation}\label{LagCoup}
{\cal L}=e\tilde{H}_{ab} \bar{\psi}\gamma^{[a}\gamma^{b]}\psi,
\end{equation}
where $\gamma^a$ is the contraction of Dirac matrices $\gamma^\mu$ with an orthonormal tetrad $e^a_\mu$, that is, $\gamma^a= e^a_\mu \gamma^\mu$,
\begin{eqnarray}\label{Hab}
\tilde{H}_{ab}&=&g^{cd}\sum_{\alpha,\beta=\pm}\sum_{l,m=1}^3\\
&&\left\{\left( M^{(\alpha,\beta,l,m)} G^{efgh} X_{ef}^{(\alpha,l)}X_{gh}^{(\beta,m)}
+N^{(\alpha,\beta,l,m)} \epsilon^{efgh}X_{ef}^{(\alpha,l)}X_{gh}^{(\beta,m)}
\right)X^{(\alpha,l)}_{c[a}X^{(\beta,m)}_{b]d}\right. \nonumber\\
&& + \left.\left(\widetilde{M}^{(\alpha,\beta,l,m)}G^{efgh} X_{ef}^{(\alpha,l)} \widetilde{X}_{gh}^{(\beta,m)}+\widetilde{N}^{(\alpha,\beta,l,m)}\epsilon^{efgh}X_{ef}^{(\alpha,l)} \widetilde{X}_{gh}^{(\beta,m)}\right)X^{(\alpha,l)}_{c[a}\widetilde{X}^{(\beta,m)}_{b]d} \right\},\nonumber
\end{eqnarray}
and
\begin{eqnarray}\label{couplingConstants}
\label{couplingConstant1} M^{(\alpha,\beta,l,m)}&=&\xi^{(\alpha,\beta,l,m)} |\lambda^{(\alpha,l)}|^{1/4}|\lambda^{(\beta,m)}|^{1/4} \left(\frac{|\lambda^{(\alpha,l)}|^{1/2}}{M_P}\right)^{c^{(\alpha,l)}}\left(\frac{|\lambda^{(\beta,m)}|^{1/2}}{M_P}\right)^{c^{(\beta,m)}},\\
\label{couplingConstant2} N^{(\alpha,\beta,l,m)}&=&\chi^{(\alpha,\beta,l,m)} |\lambda^{(\alpha,l)}|^{1/4}|\lambda^{(\beta,m)}|^{1/4} \left(\frac{|\lambda^{(\alpha,l)}|^{1/2}}{M_P}\right)^{d^{(\alpha,l)}}\left(\frac{|\lambda^{(\beta,m)}|^{1/2}}{M_P}\right)^{d^{(\beta,m)}},\\
\label{couplingConstant3} \widetilde{M}^{(\alpha,\beta,l,m)}&=&\widetilde{\xi}^{(\alpha,\beta,l,m)}|\lambda^{(\alpha,l)}|^{1/4}|\lambda^{(\beta,m)}|^{1/4}\left(\frac{|\lambda^{(\alpha,l)}|^{1/2}}{M_P}\right)^{\widetilde{c}^{(\alpha,l)}}\left(\frac{|\lambda^{(\beta,m)}|^{1/2}}{M_P}\right)^{\widetilde{c}^{(\beta,m)}},\\
\label{couplingConstant4} \widetilde{N}^{(\alpha,\beta,l,m)}&=&\widetilde{\chi}^{(\alpha,\beta,l,m)}|\lambda^{(\alpha,l)}|^{1/4}|\lambda^{(\beta,m)}|^{1/4}\left(\frac{|\lambda^{(\alpha,l)}|^{1/2}}{M_P}\right)^{\widetilde{d}^{(\alpha,l)}}\left(\frac{|\lambda^{(\beta,m)}|^{1/2}}{M_P}\right)^{\widetilde{d}^{(\beta,m)}}.
\end{eqnarray}
The free dimensionless parameters of the model are $\xi^{(\alpha,\beta,l,m)}$, $\widetilde{\xi}^{(\alpha,\beta,l,m)}$, $\chi^{(\alpha,\beta,l,m)}$, $\widetilde{\chi}^{(\alpha,\beta,l,m)}$, $c^{(\alpha,l)}$, $\widetilde{c}^{(\alpha,l)}$, $d^{(\alpha,l)}$ and $\widetilde{d}^{(\alpha,l)}$ and are subject to the restriction
\begin{equation}
c^{(\alpha,l)},\widetilde{c}^{(\alpha,l)},d^{(\alpha,l)},\widetilde{d}^{(\alpha,l)}>-1/2,
\end{equation}
to guarantee that the coupling vanishes in flat space-time regions. Although it is not evident at first sight, this is the most general field $\tilde{H}_{ab}$ that can be built respecting the restrictions discussed above \cite{Unambigous} because the terms involving two ``tilded'' eigenforms are equivalent to terms already contained in equation (\ref{Hab}). Also, we remark that the power of $M_P$ is a free parameter of the model, a peculiar feature discussed in more detail below.

Observe that the Lagrangian terms (\ref{Lagrangian}) resemble the $H_{ab}$ coupling in the Standard-Model Extension (SME) \cite{SME, SMEFields} (that is the reason why we denoted our object by $\tilde{H_{ab}}$). In this sense, the model presented here can be considered as a concrete construction of the $H_{ab}$ in the SME in terms of space-time curvature. However, we must emphasize that the two proposals are in fact of rather different nature (and that is  why  we  used the tilde to distinguish between the two). One the one hand, the fields being  considered  in  the context of the SME  represent  either  constant fields in space-time, within the purely special relativistic context, or, in the broader context provided by general relativity, the fields must correspond  to some sort of "vacuum expectation values" of tensor fields. In any case, these new background fields are taken to  be essentially constant within the space-time region of interest.  On the other hand, that conceptual difference, leads to an 
important difference regarding the experimental situation. In fact  the existing experimental bounds on these SME coefficients (see Ref. \cite{datatables}) are obtained by assuming that the fields like $H_{ab}$ can be separated into a background plus small fluctuations (which  are not taken to play a role in the test at hand), with the background fields fixed in the standard Sun-centered frame. Thus, the sidereal and orbital changes in the orientation of the laboratory with respect to the background fields produce effects with known periods, and such modulations become the focus of the observational studies. In contrast, in the model we consider here, all objects are determined locally by the space-time structure. The gravitational environment is the only source of any exotic effects and, if the gravitational 
environment in the vicinity of the experiment is unchanged, there are no modulated signals. Thus, experiments that report bounds on these SME coefficients cannot be used to establish bounds on our model unless there is a modulation of the local gravitational environment\footnote{After a second thought one can note that the gravitational effect exerted by the Sun on the Earth does vary with a sidereal frequency due to the Earth's orbit ellipticity. This observation allowed to bound the free parameters of the model \cite{Unambigous}.}. In the next section we turn our attention to the phenomenological implications of the model and the experiment that was used to test it.
 
\section{Phenomenology with the proposed interaction}\label{phen}

\noindent The model presented in this work is defined for all fermions as probing particles, however, in this section we focus on electrons. Furthermore, we use Riemann normal coordinates around the point where space-time is probed and introduce a notation where capital letters $A$, $B$, $C$, and $D$ represent pairs of antisymmetric space-time indices, taking values in roman numerals with the convention $I = 01$, $II = 02$, $III = 03$, $IV = 23$, $V = 31$, and $VI = 12$. Note that contraction with capital-letter indices and contraction with space-time indices differ by a factor $1/2$, which can be absorbed nevertheless through a rescaling of the coupling parameters. In this case the metric in $\mathcal{S}$ and the natural space-time volume $4$-form ${\epsilon_{AB}}$ are expressed as
\begin{equation}
 G_{AB} = 
 \frac{1}{2}\begin{pmatrix}
 \textbf{-1} & \quad\textbf{0} \\
 \textbf{0} & \quad\textbf{1} \\
 \end{pmatrix},
 \qquad\qquad
 {\epsilon_A}^B = 
 \begin{pmatrix}
 \textbf{0} & \quad\textbf{1} \\
 \textbf{-1} & \quad\textbf{0} \\
 \end{pmatrix},
\end{equation}
where $\textbf{0}$ and $\textbf{1}$ are the null and identity $3\times 3$ matrices, while considering only the effects of gravity in the usual linear approximation the self-adjoint maps take the generic form \cite{Yuri}
\begin{equation}
 {{(W_+)}_A}^B = 
 \begin{pmatrix}
 \textbf{A} & \quad\textbf{0} \\
 \textbf{0} & \quad\textbf{A} \\
 \end{pmatrix},
 \qquad\qquad
 {{(W_-)}_A}^B = 
 \begin{pmatrix}
 \textbf{B} & \quad\textbf{0} \\
 \textbf{0} & \quad\textbf{B} \\
 \end{pmatrix},
\end{equation}
with $\textbf{A}$ and $\textbf{B}$ real traceless symmetric $3\times 3$ matrices, as required by the symmetries of the Weyl tensor, whose components in the non-relativistic limit satisfy \cite{Unambigous}
\begin{equation}\label{weylmatrices}
 A_{ij}= \partial_i \partial_j \Phi_N,\quad\quad\quad
 B_{ij}= 2 \epsilon_{kl(i} \partial_{j)} \partial_k \Pi_l, 
\end{equation}
where $i,j,k,l$ are spacial indices running from $1$ to $3$, $\epsilon_{jkl}$ is the totally anti-symmetric tensor ($\epsilon_{123}=1$), $\Phi_N$ is the Newtonian potential of the gravitational sources and
\begin{equation}
 \Pi^i= \int \frac{p^i(t,\vec{x}')}{|\vec{x}-\vec{x}'|}d^3x',
\end{equation}
where in this last equation $p^i$ is the momentum density of the gravitational source, the arrow stands for a $3$-vector and the norm in the denominator is the one associated with an Euclidean metric. It is important to note that $\textbf{B}$ is typically suppressed by a factor $1/c$ with respect to $\textbf{A}$. Nevertheless, we cannot neglect the matrix $\textbf{B}$ because its eigenvectors are crucial for having a non-vanishing coupling term. The eigenvalue problem of ${{(W_\pm)}_A}^B$ can be solved by finding the eigenvectors $\vec{a}^{(l)}$ and $\vec{b}^{(l)}$ and eigenvalues $\alpha^{(l)}$ and $\beta^{(l)}$, such that
\begin{equation}\label{weylEigen}
 {A_i}^ja_j^{(l)} = \alpha^{(l)}a_i^{(l)}, \quad\quad\quad
 {B_i}^jb_j^{(l)} = \beta^{(l)}b_i^{(l)}.
\end{equation}
Clearly, $\lambda^{(+,l)} = \alpha^{(l)}$ and $\lambda^{(-,l)} = \beta^{(l)}$. The equations relating $X_{ab}^{(\pm,l)}$ with $a_i^{(l)}$ and $b_i^{(l)}$ can be also easily found and are given in Ref. \cite{Unambigous}.

We can simplify our calculations further if we obtain the low energy Hamiltonian associated with the Lagrangian (\ref{LagCoup}). In order to do so it is useful to exploit the similarity with the SME. In a small vicinity of a point where we want the proposed effect to be probed, ${\tilde{H}}_{ab}$ is almost constant and it can be identified with the $H_{ab}$ contained in the fermionic sector of the SME \cite{SME}. Thus, the relevant contributions to the low energy Hamiltonian of our model can be read off from the formulation of the low energy Hamiltonian for the SME \cite{KostNonRel}, yielding
\begin{equation}
 \mathcal{H} = \epsilon^{ijk}\left\{\frac{1}{2}\left[\sigma_i+\left(\vec{\sigma}\cdot\frac{\vec{p}}{m}\right)\frac{p_i}{m}\right]{\tilde{H}}_{jk} + \left(1-\frac{1}{2}\frac{p^2}{m^2}\right)\frac{p_i}{m}\sigma_j{\tilde{H}}_{0k}\right\},
\end{equation}
where $\vec{p}$ and $m$ are respectively the momentum and mass of the test particle and $\sigma_i$ are the Pauli matrices. From equation (\ref{Hab}), it can be shown that
\begin{eqnarray}
\tilde{H}_{0i} & = & g^{jk}\sum_{\alpha,\beta=\pm}\sum_{l,m=1}^3\widetilde{N}^{(\alpha,\beta,l,m)}\epsilon^{abcd}X_{ab}^{(\alpha,l)}\widetilde{X}_{cd}^{(\beta,m)} X^{(\alpha,l)}_{0[j}\widetilde{X}^{(\beta,m)}_{i]k}\nonumber \\
& = &\sum_{l,m=1}^3\left[\widetilde{N}^{(+,-,l,m)} - \widetilde{N}^{(-,+,m,l)}\right] \left[\vec{a}^{(l)}\cdot\vec{b}^{(m)}\right]\left[\vec{a}^{(l)}\times\vec{b}^{(m)}\right]_i,\\
\frac{1}{2}{\epsilon_i}^{jk}\tilde{H}_{jk} & =& {\epsilon_i}^{jk}\sum_{\alpha,\beta=\pm}\sum_{l,m=1}^3 M^{(\alpha,\beta,l,m)} G^{abcd} X_{ab}^{(\alpha,l)}X_{cd}^{(\beta,m)}X^{(\alpha,l)}_{0[j}X^{(\beta,m)}_{k]0}\nonumber \\
& =& \sum_{l,m=1}^3\left[M^{(+,-,l,m)} - M^{(-,+,m,l)}\right] \left[\vec{a}^{(l)}\cdot\vec{b}^{(m)}\right]\left[\vec{a}^{(l)}\times\vec{b}^{(m)}\right]_i,
\end{eqnarray}
where the usual notation for the Cartesian scalar and vector product is used. Therefore, it can be concluded that the expression for the non-relativistic Hamiltonian for our model is
\begin{equation}\label{HNR}
\begin{split}
\mathcal{H} = \sum_{l,m=1}^3 \left[\vec{a}^{(l)}\cdot\vec{b}^{(m)}\right]\left[\vec{a}^{(l)}\times\vec{b}^{(m)}\right]_i & \left\{\left[\sigma^i+\left(\vec{\sigma}\cdot\frac{\vec{p}}{m}\right)\frac{p_i}{m}\right]\left[M^{(+,-,l,m)} - M^{(-,+,m,l)}\right] \right. \\
& \quad \left. + \left(1-\frac{1}{2}\frac{p^2}{m^2}\right)\frac{(\vec{p}\times\vec{\sigma})^i}{m}\left[\widetilde{N}^{(+,-,l,m)} - \widetilde{N}^{(-,+,m,l)}\right]\right\}.
\end{split}
\end{equation}
Finally, in a reference frame where the probe particle is at rest, the low energy Hamiltonian takes a very simple form,
\begin{equation}\label{HNRrest}
\mathcal{H} = \sum_{l,m=1}^3 \left[M^{(+,-,l,m)} - M^{(-,+,m,l)}\right] \left[\vec{a}^{(l)}\cdot\vec{b}^{(m)}\right]\left[\vec{a}^{(l)}\times\vec{b}^{(m)}\right]\cdot\vec{\sigma}.
\end{equation}
This Hamiltonian allows to test this quantum gravity phenomenology proposal by establishing a bridge with experimental scenarios involving fermions at low energies. We remark that, although the model is covariant, the expressions for $\tilde{H}_{0i}$ and $\tilde{H}_{jk}$ found above do not seem to be covariantly related. This is only because we are presenting approximated expressions where the $O(c^{-2})$ terms are neglected and by the fact that the Lorentz transformation that mixes $\tilde{H}_{0i}$ with $\tilde{H}_{jk}$ involves an extra factor $c^{-1}$.

The appearance of the Pauli matrices in the Hamiltonian (\ref{HNR}) suggests that only polarized matter is sensible to the proposed coupling. This rendered a feasible experimental scenario for testing this model difficult to find, since, on one hand, most of bulk matter as immediately found in nature is not polarized, and, on the other hand, polarized matter possesses a magnetic moment whose magnetic effects interfere with the phenomenon that is being investigated. The E\"ot-Wash group had developed an experimental technique consisting of a continuously rotating torsion balance which involves approximately $10^{23}$ horizontally polarized electrons possessing negligible magnetic moment \cite{Heckel}. Therefore, some of us were impelled to take advantage of this existing technology and propose a compatible experimental setting \cite{searchingBS} for testing the model here presented. Later, the E\"ot-Wash group performed the experiment suggested in Ref. \cite{searchingBS} which resulted in bounds on some of 
the free parameters of the model, corresponding to the features the experiment was sensitive to. This is the subject of the next section.

\section{The experiment} \label{exp}

\noindent The experimental setting essentially consists of a torsion pendulum containing a number of (0.96$\pm$0.17)$\times 10^{23}$ horizontally polarized electrons with negligible magnetic moment, subject to the gravitational field of controlled sources and the Earth and to the frame dragging effect due to the Earth's angular momentum. The controlled gravitational sources produce a gravitational field which is equivalent to the field generated by two identical point masses of $38.4 \ \rm{kg}$ located symmetrically with respect to the torsion pendulum on a horizontal plane $4.0 \ \rm{cm}$ above the center of the pendulum and at a distance of $28.0 \ \rm{cm}$ from the torsion pendulum \cite{terrano}.

The gravitational configuration described above determines the Weyl tensor. As we mentioned above, we are working with normal Riemann coordinates associated with the point where the probe is located. Considering a remaining freedom in choosing coordinate systems rotated one from the other, we choose coordinates such that the $z$ coordinate points up, the $y$ coordinate points south and the $x$ coordinate points to the east. The matrices $\mathbf{A}$ and $\mathbf{B}$ are calculated using equations (\ref{weylmatrices}). Assuming that the laboratory sources of mass $m$ are located at $(d\cos\theta,d\sin\theta,z_0)$ and $(-d\cos\theta,-d\sin\theta,z_0)$ in the specified coordinates, the matrices corresponding to these sources are
\begin{equation}
 \mathbf{A}_L = \frac{md^2}{(d^2+z_0^2)^{5/2}}
 \begin{pmatrix}
 1+3\cos(2\theta)-2z_0^2/d^2	& 3\sin(2\theta)			& 0		 \\
 3\sin(2\theta)			& 1-3\cos(2\theta)-2z_0^2/d^2		& 0		 \\
 0					& 0					& -2+4z_0^2/d^2 \\
 \end{pmatrix}
\end{equation}
and $\mathbf{B}_L = \mathbf{0}$ because this matrix depends on the momentum of the gravitational sources which, in this case, is zero. The effect of the Earth's gravitational source is two-fold, its mass distribution yields a gravitational potential and its rotation produces a frame dragging effect. We compute the Weyl tensor accounting for these aspects in a different manner from the use of equations (\ref{weylmatrices}). Instead, we consider the linearized Weyl tensor corresponding to the Kerr metric, in the reference frame instantaneously at rest with respect to the probing particle. We obtain therefore
\begin{equation}
 \mathbf{A}_\oplus = \frac{M}{R^3}
 \begin{pmatrix}
 -1 & 0 & 0 \\
 0 & -1 & 0 \\
 0 & 0 & 2 \\
 \end{pmatrix},
 \quad\quad \mathbf{B}_\oplus = \frac{3J}{R^4}
 \begin{pmatrix}
 \cos\theta_L		& 0			& 0			\\
 0			& \cos\theta_L		& -\sin\theta_L	\\
 0			& -\sin\theta_L	& -2\cos\theta_L	\\
 \end{pmatrix},
\end{equation}
where $\theta_L$ represents the co-latitude of the laboratory, $\oplus$ is used to denote the Earth, and $R$, $M$ and $J$ represent the Earth's radius, mass and angular momentum, respectively. In the linearized regime, the matrices accounting for all gravitational sources are given by
\begin{equation}
 \mathbf{A} = \mathbf{A}_L + \mathbf{A}_\oplus, \quad\quad \mathbf{B} = \mathbf{B}_\oplus.
\end{equation}
From these, it is straightforward to compute both their orthonormal eigenvectors and eigenvalues as defined by (\ref{weylEigen}). The eigenvalues for $\mathbf{A}$ are 
\begin{eqnarray}
\alpha^{(1)} & = & \frac{2M}{R^3}-\frac{2m \left(d^2-2 z_0^2\right)}{\left(d^2+z_0^2\right)^{5/2}},\\
\alpha^{(2)} & = & -\frac{M}{R^3}-\frac{2m}{\left(d^2+z_0^2\right)^{3/2}},\\
\alpha^{(3)} & = & -\frac{M}{R^3}+\frac{2m(2d^2 -z_0^2)}{\left(d^2+z_0^2\right)^{5/2}},
\end{eqnarray}
while those of $\mathbf{B}$ are
\begin{eqnarray}
\beta^{(1)} & = & \frac{3J}{R^4} \cos\theta_L,\\
\beta^{(2)} & = & \frac{3J}{2R^4}\left(-\cos\theta_L -\sqrt{5\cos^2\theta_L+4}\right), \\
\beta^{(3)} & = & \frac{3J}{2R^4}\left(-\cos\theta_L +\sqrt{5\cos^2\theta_L+4}\right),
\end{eqnarray}
with corresponding orthonormal eigenvectors
\begin{eqnarray}
\vec{a}^{(1)} & = & (0,0,1),\\
\vec{a}^{(2)} & = &|\cos\theta| (-\tan \theta ,1,0),\\
\vec{a}^{(3)} & = & |\sin\theta|(\cot\theta ,1,0), 
\end{eqnarray}
and
\begin{eqnarray}
\vec{b}^{(1)} & = & (1,0,0),\\
\vec{b}^{(2)} & = &(0,-3 \cos \theta_L+S,2 \sin\theta_L)/N_+,\\
\vec{b}^{(3)} & = &(0,-3 \cos \theta_L-S,2 \sin\theta_L)/N_- , 
\end{eqnarray}
respectively, where
\begin{equation}
S=\sqrt{5 \cos^2\theta_L+4},\qquad N_{\pm}=\sqrt{10 \cos^2\theta_L \mp 6S\cos\theta_L+8}.
\end{equation}
Observe that in this derivation we assume $\sin \theta_L>0$, thus, the expressions for $\vec{b}^{(2)}$ and $\vec{b}^{(3)}$ are valid in any place on Earth excepting its poles. The experimental configuration determines entirely the values of the eigenvalues $\alpha^{(i)}$ and $\beta^{(i)}$. Considering that the experiment was performed in a laboratory at a co-latitude of $\theta_L = 42.3^\circ$, the eigenvalues are listed in Table \ref{eigenValues}.

\begin{table}[h]
\centering
\begin{tabular}{lll}
 \hline
 i & $\sqrt{\alpha^{(i)}}$ (eV) & $\sqrt{\beta^{(i)}}$ (eV)\\
 \hline
 1 \quad & $1.1\times10^{-18}$ & $9.6\times10^{-22}$ \\
 2 \quad & $8.7\times10^{-19}$ & $1.4\times10^{-21}$ \\
 3 \quad & $6.9\times10^{-19}$ & $1.1\times10^{-21}$ \\
 \hline
\end{tabular}
\caption{Eigenvalues $\alpha^{(i)}$ and $\beta^{(i)}$.}
\label{eigenValues}
\end{table}

Thanks to the fact that both $\mathbf{A}$ and $\mathbf{B}$ are non-null, the gravitational fields to which the electrons are subject leads to a non-null low energy Hamiltonian that, according to equation (\ref{HNRrest}), in the laboratory reference frame takes the form
\begin{equation}
\begin{split}
\mathcal{H}^e & = N\sum_{l,m=1}^3\Delta\xi^{(l,m)}|\alpha^{(l)}|^{1/4}|\beta^{(m)}|^{1/4}\left(\frac{|\alpha^{(l)}|^{1/2}}{M_P}\right)^{c^{(+,l)}} \left(\frac{|\beta^{(m)}|^{1/2}}{M_P}\right)^{c^{(-,m)}} \\ 
& \quad\qquad\qquad \left[\vec{a}^{(l)}\cdot\vec{b}^{(m)}\right]\left[\vec{a}^{(l)}\times\vec{b}^{(m)}\right]\cdot\vec{\sigma},
\end{split}
\end{equation}
where $\Delta\xi^{(l,m)}=\xi^{(+,-,l,m)}-\xi^{(-,+,m,l)}$ is a free parameter of the model and the definition of $M^{(\alpha,\beta,l,m)}$, equation (\ref{couplingConstant1}), is employed. The expectation value of the Hamiltonian for the $N$ electrons, each with spin pointing at the direction $\hat{n} = (\cos\phi,\sin\phi,0)$, is
\begin{equation}\label{elEn}
\begin{split}
E^e = \langle\mathcal{H}^e\rangle & = N\sum_{l,m=1}^3\Delta\xi^{(l,m)}|\alpha^{(l)}|^{1/4}|\beta^{(m)}|^{1/4}\left(\frac{|\alpha^{(l)}|^{1/2}}{M_P}\right)^{c^{(+,l)}}\left(\frac{|\beta^{(m)}|^{1/2}}{M_P}\right)^{c^{(-,m)}} \\ 
& \quad\qquad\qquad \left[\vec{a}^{(l)}\cdot\vec{b}^{(m)}\right]\left[\vec{a}^{(l)}\times\vec{b}^{(m)}\right]\cdot\hat{n}.
\end{split}
\end{equation}
Note that we can write the total energy of the torsion pendulum as $E^e = -N\hat{n}\cdot\vec{w}$, for a vector $\vec{w}$ that can be read off from equation (\ref{elEn}). Thus, the torque exerted on the torsion pendulum is $\vec{T} = N\hat{n}\times\vec{w}$, implying that its $z$ component satisfies
\begin{equation}\label{torque}
\begin{split}
T_z & = N\sum_{l,m=1}^3\Delta\xi^{(l,m)}|\alpha^{(l)}|^{1/4}|\beta^{(m)}|^{1/4}\left(\frac{|\alpha^{(l)}|^{1/2}}{M_P}\right)^{c^{(+,l)}}\left(\frac{|\beta^{(m)}|^{1/2}}{M_P}\right)^{c^{(-,m)}} \\
& \quad\times \left(f_y^{(l,m)}\cos\phi-f_x^{(l,m)}\sin\phi\right),
\end{split}
\end{equation}
where
\begin{equation}\label{fCoeffs}
 \begin{split}
 f_x^{(1,2)} & = -f_x^{(1,3)} = -I \\
 f_x^{(2,2)} & = -f_x^{(2,3)} = I\cos^2\theta \\
 f_x^{(3,2)} & = -f_x^{(3,3)} = I\sin^2\theta \\
 f_y^{(2,2)} & = f_y^{(3,3)} = -f_y^{(2,3)} = -f_y^{(3,2)} = I\cos\theta\sin\theta,
 \end{split}
\end{equation}
and the rest of the functions vanish. In the last expressions, $I = \sin\theta_L/\sqrt{5\cos^2\theta_L+4}$. As can be seen from the form of the coefficients (\ref{fCoeffs}), the torque exerted on the torsion pendulum depends on $\theta$, which can be controlled in the laboratory by rotating the sources. 

The detection of the torque (\ref{torque}) would be the experimental signal arising from the proposed model. However, the E\"ot-Wash group detected no torque \cite{terrano} leading to the following bounds,
\begin{equation}
 \begin{split}
 & \left|9.19\Delta\xi^{(2,2)}\left(\frac{|\alpha^{(2)}|^{1/2}}{M_P}\right)^{c^{(+,2)}}\left(\frac{|\beta^{(2)}|^{1/2}}{M_P}\right)^{c^{(-,2)}}\right. \\
 & \left. - 7.94\Delta\xi^{(2,3)}\left(\frac{|\alpha^{(2)}|^{1/2}}{M_P}\right)^{c^{(+,2)}}\left(\frac{|\beta^{(3)}|^{1/2}}{M_P}\right)^{c^{(-,3)}}\right. \\ 
 & \left. - 8.17\Delta\xi^{(3,2)}\left(\frac{|\alpha^{(3)}|^{1/2}}{M_P}\right)^{c^{(+,3)}}\left(\frac{|\beta^{(2)}|^{1/2}}{M_P}\right)^{c^{(-,2)}}\right. \\ 
 & \left. + 7.06\Delta\xi^{(3,3)}\left(\frac{|\alpha^{(3)}|^{1/2}}{M_P}\right)^{c^{(+,3)}}\left(\frac{|\beta^{(3)}|^{1/2}}{M_P}\right)^{c^{(-,3)}}\right|<3.7.
 \end{split}
\end{equation}
By assuming that there are no cancellations among terms in the l.h.s., each term is bounded by
\begin{eqnarray}
 9.19\left|\Delta\xi^{(2,2)}\right|\left(\frac{|\alpha^{(2)}|^{1/2}}{M_P}\right)^{c^{(+,2)}}\left(\frac{|\beta^{(2)}|^{1/2}}{M_P}\right)^{c^{(-,2)}} & < 3.7 \\
 7.94\left|\Delta\xi^{(2,3)}\right|\left(\frac{|\alpha^{(2)}|^{1/2}}{M_P}\right)^{c^{(+,2)}}\left(\frac{|\beta^{(3)}|^{1/2}}{M_P}\right)^{c^{(-,3)}} & < 3.7 \\ 
 8.17\left|\Delta\xi^{(3,2)}\right|\left(\frac{|\alpha^{(3)}|^{1/2}}{M_P}\right)^{c^{(+,3)}}\left(\frac{|\beta^{(2)}|^{1/2}}{M_P}\right)^{c^{(-,2)}} & < 3.7 \\ 
 7.06\left|\Delta\xi^{(3,3)}\right|\left(\frac{|\alpha^{(3)}|^{1/2}}{M_P}\right)^{c^{(+,3)}}\left(\frac{|\beta^{(3)}|^{1/2}}{M_P}\right)^{c^{(-,3)}} & <3.7.
\end{eqnarray}
Each of these inequalities involves three different free parameters, $\Delta\xi^{(l,m)}$, $c^{(+,l)}$, and $c^{(-,m)}$ and comprise a specific region in parameter space where it is satisfied. A way to visualize these inequalities is as tridimensional projections of the entire parameter space, as shown in Fig. \ref{fig:validParameters1}. Our hope is that in the near future more experiments will be done to test more precisely the predictions of the model and to probe different parts of the parameter space, including sectors associated with other fermions. Also, that the bounds presented in Fig. \ref{fig:validParameters1} could be used to test candidate theories of quantum gravity.

\setcounter{table}{0}
\renewcommand{\tablename}{FIGURE} 

\begin{table}[h]
\centering
\begin{tabular}{cc}
\includegraphics[width=0.48\textwidth]{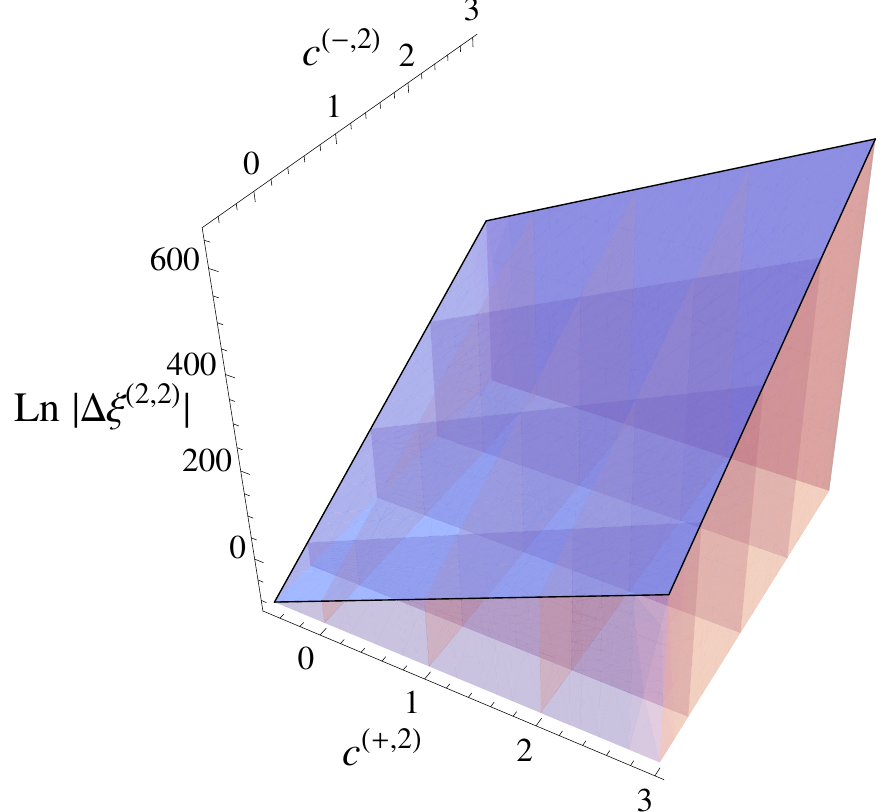} & \includegraphics[width=0.48\textwidth]{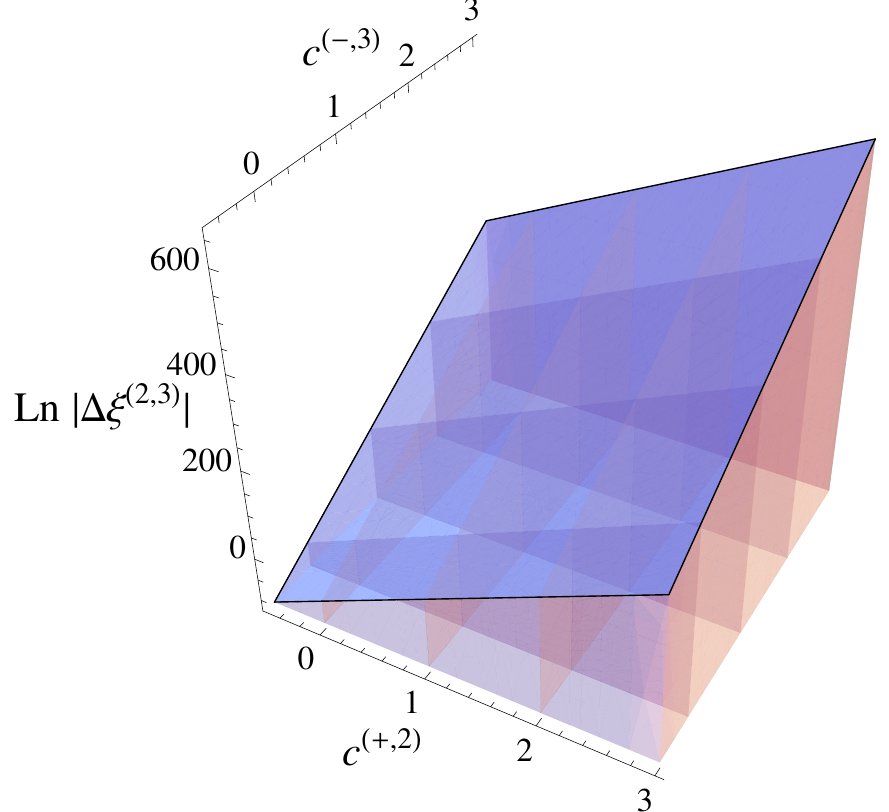} \\
\includegraphics[width=0.48\textwidth]{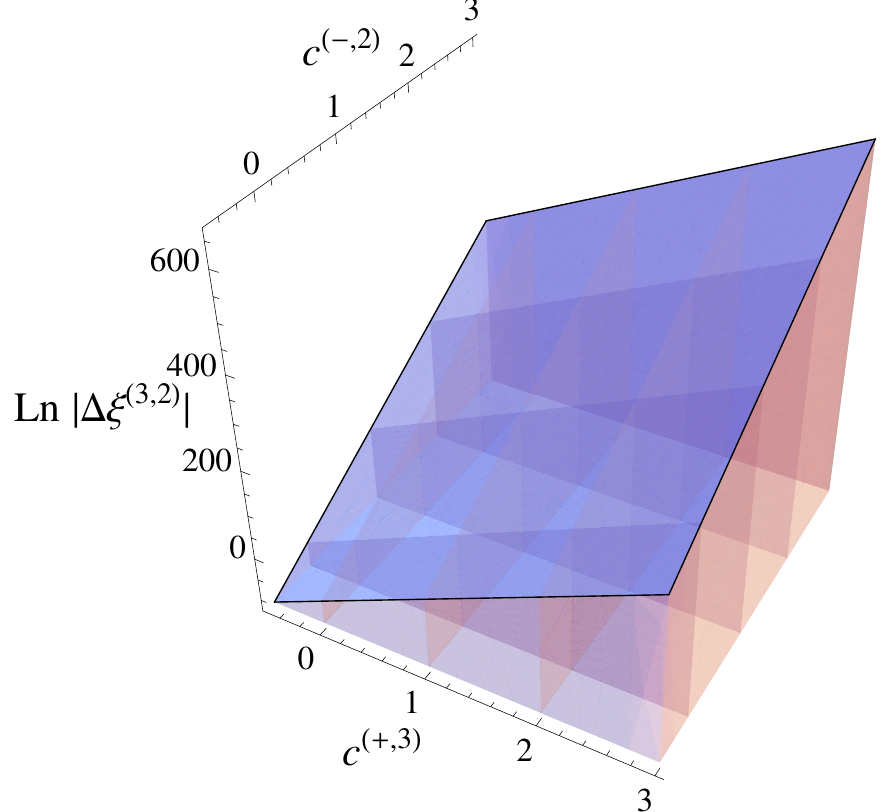} & \includegraphics[width=0.48\textwidth]{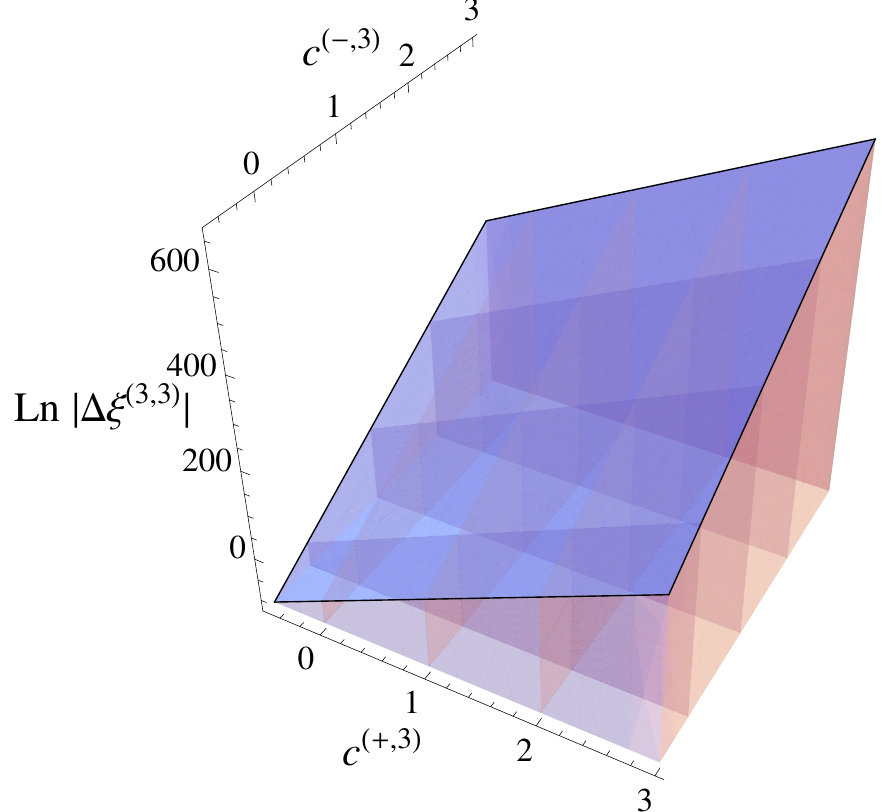} \\
\end{tabular}
\caption{Sections of the parameter space for the model presented in this work. 
The empty parts of the graphs correspond to experimentally discarded regions.}
\label{fig:validParameters1}
\end{table}

\setcounter{table}{1}
\renewcommand{\tablename}{TABLE} 


\section{Counterintuitive features of the model} \label{counterintuitive}

\noindent A careful inspection to the model presented here can raise various concerns. As a first remark, the model is not well-defined for all space-times because the eigenvalue equation (\ref{eigen map}) can be further degenerated (as in any space-time which is not type I according to Petrov's classification \cite{Petrov,HallSCSGR}). In addition, what is supposed to codify the effects of the granular space-time structure is described in terms of unconventional objects which so far are not known to play a role in the description of space-time. Moreover, the model incorporates some particular behavior under simple space-time symmetries: for instance, even if one accepts that the model invokes some coupling between angular momentum of the source (Earth) and the fermion spin, one can expect that upon reversal of direction of the Earth's rotation (keeping all other things unchanged), the effects would reverse sign as well. In our model this does not occur, as can be easily observed from the quadratic dependence 
of the 
non-relativistic Hamiltonian (\ref{HNR}) in $\vec{a}^{(l)}$ and $\vec{b}^{(l)}$ and the even parity of their associated eigenvalues, which are determined by the gravitational sources. Also, the free parameters appearing as exponents in (\ref{couplingConstant1}-\ref{couplingConstant4}) are highly not standard; normally one would expect such exponents to be integers. In fact, non-integer exponents lead to non-analyticity of the expressions.

This initial appearance of strangeness and unnaturalness should be reconsidered after recalling that according to the point of view adopted in this work, the geometry must be considered as emerging from a fundamental quantum description which does not characterize the fundamental degrees of freedom in terms of the space-time metric that one employs in classical general relativity. In this vein, we must analyze other phenomena known as the effective and macroscopic description of some more fundamental substrate for which we have a good theoretical understanding, and consider whether in those cases counterintuitive features can arise. Therefore, we resort back to the analogy with the hydrodynamical description of a liquid using the N-S equation, and consider cases where unusual behavior is not easily codified in the natural variables associated with such equations. The fact that the model is not well defined for every physical situation, is clearly analogous to the fact that 
the N-S equation cannot describe any conceivable fluid in all conceivable situations: it should be clear, for instance, that a wave on the seashore, after breaking with the subsequent production of foam, will not have a simple hydrodynamical description. 
 
Concerning the counterintuitive behavior of the effect under a change in the sense of rotation of the Earth, we next analyze if something similar can happen in other contexts. Let's consider a nematic liquid made out of long molecules (think of them as un-oriented roads, one extreme being the same as the other). Now let us examine how the molecules align if the fluid is placed between two solid parallel plates moving at different speeds in one direction, say $x$ (with that direction tangent to the planes). It seems clear that the molecules tend to align in that same direction because other alignments increase the collisions between molecules, and those collisions tend to reorient them. Imagine we now proceed to investigate the fluid employing polarized radiation. Assuming that the electrons within the long molecules move rather freely, then, it is clear that the interaction with radiation reduces the polarization of the radiation in the $x$ direction, unlike in perpendicular directions. If we 
now consider the same liquid, but in a different setting, where there are vortices associated with rotation (for instance, a liquid draining in a sink), the molecules will again align along the direction of flow of the liquid (again due to the difference in velocities), and this will again affect the liquid's interaction with polarized light in a polarization dependent way. However, it is clear that the effect on the light's polarization is the same whether the liquid is rotating in one sense or in the opposite one. Thus, the effect is approximately proportional to the angular momentum of the sinking fluid, but it does not change sign when the direction of the rotation changes. This is an example of an unusual behavior under a change of symmetry that is not unexpected in such situation and seems very similar to what occurs in our model regarding the behavior of the effect under the inversion of the Earth's rotation. 

Note further that our model incorporates the possibility of parity chance (P) and time reversal (T) violation through the appearance of the volume $4$-form in the construction of section II. The exact dependence of the model on the space-time orientation is not easy to see because the model involves various manipulations and selections of the volume form. The point is that we see no problem in the fact that there are Lagrangian terms that are T odd and P even. In fact, it is a possibility that the fundamental space-time structure could involve, in addition to the metric, spatial and time orientations. Moreover, the fact that gravity could violate T and/or P at the fundamental level is somehow appealing because gravity, as described by Einstein's theory, dramatically violates the pattern shown by the other interactions, namely, that weaker interactions respect less symmetries. 

Finally we note that the non-analyticity arising from non-integer exponents is not unnatural in effective descriptions either. We recall for instance the large range of phenomena which are described by expressions associated with critical exponents (see Ref. \cite{Zee}).

\section{Conclusions}\label{discussion}

\noindent The basic constituents of our world, as far as we understand them now, are the matter fields, described by the standard model of particle physics augmented to incorporate the masses of neutrinos and the inflaton, the still mysterious dark matter and dark energy, and the gravitational sector. Of these, the standard fields including the electroweak and strong interactions, as well as the quarks and leptons, and the inflaton, seem to fit quite nicely into the framework of quantum field theory. Dark matter is likely to be described by some other fields with a similar structure as that given for the ordinary matter, perhaps involving novel aspects like super-symmetry, and dark energy seems to be most economically described by a cosmological constant (although certain naturalness issues and questions of coincidence are still outstanding). However, the component of our world which seems hardest to fit within the general paradigms offered by quantum theory is gravitation.

There exists a very extensive literature on this subject and we cannot even attempt to describe the problems, either technical or conceptual, found in this road. However, it seems quite clear that conceptually there is room for vast differences from the usual cases (like Maxwell theory or the standard electroweak model) arising when considering the incorporation of the quantum aspects of nature in the gravitational context. According to general relativity, gravitation reflects the structure of space-time itself, whereas quantum theory seems to fit most easily in contexts where this structure is a given one. That is, quantum states are associated with objects that \textit{live} in space-times. For instance, the standard Schr\"odinger equation specifies the time evolution of a system, the quantum states of fields characterize the system in connection to algebras of observables associated with predetermined space-time regions, and so forth. It is clear that major conceptual modifications are needed if we want 
to describe the space-time itself in a quantum language. 

This issue appears in various guises in the different approaches existing nowadays to quantum gravity, most conspicuously as the problem of time which afflicts all attempts to deal with the subject following a canonical approach and it corresponds to lingering features of the fundamental time-less (and possibly space-less \cite{Posets-Sorkin}) theory of quantum gravity. If that is the case, the emergence of space-time itself would be tied to the incorporation of such effective quantum description of matter fields living on space-time, and evolving approximately according to standard quantum field theory on curved spaces (with some small deviations which might conceivable lead to rather exotic and yet undetected types of effects). In that context it seems clear that space-time itself would be nothing but an effective description of the underlying quantum gravity reality. Ideas of this sort regarding emergent gravity have been indeed considered previously, for instance in Refs. \cite{Ashtekar:2001, Jacobson1}. 
This suggests that in the context where we consider space-time granularity, the metric must be regarded as an approximate phenomenological description, and Einstein's equations, a macroscopic description of the aggregate behavior of the fundamental components.

Having this picture in mind we raised the question of how a granular space-time structure could become manifest. In contrast with most phenomenological models of quantum gravity and motivated by the study of radiative corrections in  space-times involving Lorentz violating granularity, we take the point of view that space-time, at microscopic scales, respects Lorentz invariance. Considering this hypothesis seriously, we conclude that such granularity could become manifest through nonminimal couplings of space-time curvature and matter fields. We reviewed the proposal of  developing  a model realizing these ideas and briefly described the experiment where such ideas were first tested. We have here analyzed the results in detail and obtained the corresponding  bounds on the free parameters of the model. Our hope is that these experimental bounds, and those that might arise from future and more precise experiments, can be important guidelines in the development of proposals for a quantum theory of space-time. 
In this regard we can only note the important role that the experimental tests of the universality of free fall played in the emergence of the Einsteinian ideas behind the equivalence principle, and subsequently on the development of the general theory of relativity.

\acknowledgments

\noindent We acknowledge very useful discussions with E.G. Adelberger, C. Chryssomalakos and V.A. Kosteleck\'y. This work was supported in part by the CONACYT grant 101712 and PAPIIT-UNAM IN107412. YB was supported in part by the Department of Energy under grant number DE-FG02-91ER40661 and by the Indiana University Center for Spacetime Symmetries.


\begin{thebibliography}{99}

\bibitem{Kosteleckyinitial} V.A. Kosteleck\'y and S. Samuel, ``Spontaneous breaking of Lorentz symmetry in string theory,'' \textit{Phys. Rev.} \textbf{D 39}, 683 (1989).

\bibitem{A Camellia}
G. Amelino-Camelia, J.R. Ellis, N.E. Mavromatos, D.V. Nanopoulos and S. Sarkar, ``Potential Sensitivity of Gamma-Ray Burster Observations to Wave Dispersion in Vacuo,'' \textit{Nature} \textbf{393}, 763 (1998); [arXiv:astro-ph/9712103].

\bibitem{Kozameh-Gleiser}
R.J. Gleiser and C.N. Kozameh, ``Astrophysical limits on quantum gravity motivated birefringence,'' \textit{Phys. Rev.} \textbf{D 64}, 083007 (2001).

\bibitem{Jacobson}
T. Jacobson, S. Liberati and D. Mattingly, ``Lorentz violation and Crab synchrotron emission: A new constraint far beyond the Planck scale,'' \textit{Nature} \textbf{424}, 1019 (2003). 
 
\bibitem{Mattingly}
D. Mattingly, ``Modern tests of Lorentz invariance,'' \textit{Living Rev. Rel.} \textbf{8}, 5 (2005); [arXiv:gr-qc/0502097].

\bibitem{Gambini}
R. Gambini and J. Pullin, ``Nonstandard optics from quantum spacetime,'' \textit{Phys. Rev.} \textbf{D 59}, 124021 (1999); [arXiv:gr-qc/9809038]. 

\bibitem{Alfaro1999}
J. Alfaro, H. A. Morales-T{\'e}cotl and L. F. Urrutia, ``Quantum gravity corrections to neutrino propagation,'' \textit{Phys. Rev. Lett.} \textbf{84}, 2318 (2000); [arXiv:gr-qc/9909079].

\bibitem{Morales2002}
J. Alfaro, H.A. Morales-T{\'e}cotl and L.F. Urrutia, ``Loop quantum gravity and light propagation'', \textit{Phys. Rev.} \textbf{D 65}, 103509 (2002).

\bibitem{Alfaro2005}
J. Alfaro, ``Quantum gravity and Lorentz invariance violation in the standard model,'' \textit{Phys. Rev. Lett.} \textbf{94}, 221302 (2005); [arXiv:hep-th/0412295].

\bibitem{Alfaro2005-2} 
J. Alfaro, ``Quantum gravity induced Lorentz invariance violation in the standard model: hadrons,'' \textit{Phys. Rev.} \textbf{D 72}, 024027 (2005); [arXiv:hep-th/0505228].

\bibitem{Urrutia:2006}
L. F. Urrutia, ``Corrections to Flat-Space Particle Dynamics Arising from Space Granularity,'' in \textit{Special Relativity: Lecture Notes in Physics} eds. J. Ehlers and C. L\"ammerzahl (Springer, Berlin Heidelberg, 2006); [arXiv:hep-ph/0506260].

\bibitem{KosteleckyPotting}
V. A. Kosteleck\'y and R. Potting, ``CPT, strings, and meson factories'', \textit{Phys. Rev} \textbf{D 51}, 3923 (1995); [arXiv:hep-ph/9501341].

\bibitem{Ellis2000}
J. R. Ellis, N. E. Mavromatos and D. V. Nanopoulos, ``Quantum-gravitational diffusion and stochastic fluctuations in the velocity of light,'' \textit{Gen. Rel. Grav.} \textbf{32}, 127 (2000); [arXiv:gr-qc/9904068].

\bibitem{Ellis2002}
J. R. Ellis, N. E. Mavromatos and D. V. Nanopoulos, ``A microscopic recoil model for light-cone fluctuations in quantum gravity,'' \textit{Phys. Rev.} \textbf{D 61}, 027503 (1999); [arXiv:gr-qc/9906029].

\bibitem{HD} T.E. Chupp, R.J. Hoare, R.A. Loveman, E.R. Oteiza, J.M. Richardson, M.E. Wagshul and A.K. Thompson, ``Results of a new test of local Lorentz invariance: A search for mass anisotropy in ${}^{21}$Ne'', \textit{Phys. Rev. Lett.} \textbf{63}, 1541 (1989).

\bibitem{HD Bounds}
D. Sudarsky, L. Urrutia and H. Vucetich, ``Observational Bounds on Quantum Gravity Signals using Existing Data,'' \textit{Phys. Rev. Lett.} \textbf{89}, 231301 (2002).
D. Sudarsky, L. Urrutia and H. Vucetich, ``Bounds on Stringy Quantum Gravity from Low Energy Existing Data,'' \textit{Phys. Rev.} \textbf{D 68}, 024010 (2003).


\bibitem{Collins2004}
J. Collins, A. Perez, D. Sudarsky, L. Urrutia and H. Vucetich, ``Lorentz invariance in Quantum Gravity: A new fine tuning problem?," \textit{Phys. Rev. Lett.} \textbf{93}, 191301 (2004).

\bibitem{Gambini-Pullin} R. Gambini R, S. Rastgoo and J. Pullin, ``Small Lorentz violations in quantum gravity: do they lead to unacceptably large effects?'' \textit{Class. Quantum Grav.} \textbf{28}, 155005 (2011); [arXiv:1106.1417].

\bibitem{Rovelli}
C. Rovelli and S. Speziale,``Reconcile Planck-scale discreteness and the Lorentz-Fitzgerald contraction,'' \textit{Phys. Rev.} \textbf{D 67}, 064019 (2003); [arXiv:gr-qc/0205108].

\bibitem{Polshinski2011}
J. Polchinski, ``Comment on small Lorentz violations in quantum gravity: Do they lead to unacceptably large effects?,'' [arXiv:1106.6346]

\bibitem{Coleman}
S.R. Coleman and S.L. Glashow, ``High-energy tests of Lorentz invariance,'' \textit{Phys. Rev.} \textbf{D 59}, 116008 (1999); [arXiv:hep-ph/9812418].

\bibitem{Posets-Sorkin} R.D. Sorkin, ``Forks in the Road, on the Way to Quantum Gravity'', \textit{Int. J. Theor. Phys.} \textbf{36}, 2759 (1997); [arXiv:gr-qc/9706002].

\bibitem{Bekenstein:1973ur}
J. D. Bekenstein, ``Black holes and entropy,'' \textit{Phys. Rev.} \textbf{D 7}, 2333 (1973).

\bibitem{Jacobson:1995ab}
T. Jacobson, ``Thermodynamics of spacetime: The Einstein equation of state,'' \textit{Phys. Rev. Lett.} \textbf{75}, 1260 (1995); [arXiv:gr-qc/9504004].

\bibitem{Volovik2003}
G. E. Volovik, ``The Universe in a Helium Droplet,'' (Oxford University Press, USA, 2003), page 536.

\bibitem{Padmanabhan:2003gd}
T. Padmanabhan, ``Gravity and the thermodynamics of horizons,'' \textit{Phys. Rept.} \textbf{406}, 49 (2005); [arXiv:gr-qc/0311036].
 
\bibitem{Hu:2005ub}
B.L. Hu, ``Can spacetime be a condensate?,'' \textit{Int. J. Theor. Phys.} \textbf{44}, 1785 (2005); [arXiv:gr-qc/0503067].

\bibitem{WdW} B.S. DeWitt, ``Quantum Theory of Gravity. I. The Canonical Theory," \textit{Phys. Rev.} \textbf{160}, 1113 (1967).

\bibitem{RovelliLQG} C. Rovelli, \textit{Quantum Gravity}, (Cambridge University Press, 2004).

\bibitem{Isham:1992ms}
J.A. Wheeler in \textit{Battelle Reencontres} eds. C. DeWitt and J.A. Wheeler (Benjamin, New York,1968).
C.J. Isham, ``Canonical Quantum Gravity and the Problem of Time", \textbf{GIFT Seminar}, 0157228 (1992); [arXiv:qr-qc/9210011].
 
\bibitem{Pullin:2004}
R. Gambini, R.A. Porto and J. Pullin, ``Realistic Clocks, Universal Decoherence and the Black Hole Information Paradox," \textit{Phys. Rev. Lett.} {\bf 93}, 240401 (2004); [arXiv:hep-th/0406260].
R. Gambini, R.A. Porto and J. Pullin, ``Fundamental decoherence from relational time in discrete quantum gravity: Galilean covariance," \textit{Phys. Rev.} \textbf{D 70}, 124001 (2004); [arXiv:gr-qc/0408050].

\bibitem{Emerging} See for instance:
T. Jacobson, ``Thermodynamics of space-time: The Einstein equation of state,'' \textit{Phys. Rev. Lett.} {\bf 75}, 1260 (1995); [arXiv:gr-qc/9504004].
T.~Padmanabhan, ``Gravity and the thermodynamics of horizons,'' \textit{Phys. Rept.} {\bf 406}, 49 (2005); [arXiv:gr-qc/0311036].
B.~L.~Hu, ``Can spacetime be a condensate?,'' \textit{Int. J. Theor. Phys.} {\bf 44}, 1785 (2005); [arXiv:gr-qc/0503067]. 
 
\bibitem{Wald} R.M. Wald, \textit{General Relativity} (University of Chicago Press, 1984).

\bibitem{Corichi}
A. Corichi and D. Sudarsky, ``Towards a New Approach to Quantum Gravity Phenomenology,'' \textit{Int. J. Mod. Phys.} \textbf{D 14}, 1685 (2005).

\bibitem{Yuri}
Y. Bonder and D. Sudarsky, ``Quantum Gravity Phenomenology without Lorentz Invariance Violation: a detailed proposal,'' \textit{Class. Quantum Grav.} \textbf{25}, 105017 (2008); [arXiv:0709.0551].

\bibitem{Unambigous} Y. Bonder and D. Sudarsky, ``Unambiguous Quantum Gravity Phenomenology Respecting Lorentz Symmetry'', \textit{Rept. Math. Phys.} \textbf{64} 169-184 (2009); [arXiv:0811.1229].

\bibitem{HallSCSGR} G.S. Hall, \textit{Symmetries and Curvature Structure in General Relativity}, chapter 7, (World Scientific, 2004).

\bibitem{SME}
D. Colladay and V.A. Kosteleck\'y, ``Lorentz-violating extension of the standard model'', \textit{Phys. Rev.} \textbf{D 58}, 116002 (1998).

\bibitem{SMEFields}
V.A. Kosteleck\'y, ``Gravity, Lorentz violation, and the standard model'', \textit{Phys. Rev.} \textbf{D 69}, 105009 (2004); [arXiv:hep-th/0312310]. 

\bibitem{datatables}
V.A. Kosteleck\'y and N. Russell, ``Data tables for Lorentz and $CPT$ violation'', \textit{Rev. Mod. Phys.} \textbf{83}, 11 (2011); [arXiv:0801.0287v6].

\bibitem{KostNonRel}
V.A. Kosteleck\'y and C.D Lane, ``Nonrelativistic quantum Hamiltonian for Lorentz violation'', \textit{J. Math. Phys.} \textbf{40}, 6245 (1999).

\bibitem{Heckel} B.R. Heckel, C.E. Cramer, T.S. Cook, S. Schlamminger, E.G. Adelberger, and U. Schmidt, ``New CP-Violation and Preferred-Frame Tests with Polarized Electrons'', \textit{Phys. Rev. Lett.} \textbf{97}, 021603 (2006); [arXiv:hep-ph/0606218].

\bibitem{searchingBS} Y. Bonder and D. Sudarsky, ``Searching for spacetime granularity: Analyzing a concrete experimental setup'', \textit{AIP Conf.Proc.} \textbf{1256} 157-163 (2010); [arXiv:1003.5245].

\bibitem{terrano} W.A. Terrano, B.R. Heckel and E.G. Adelberger, ``Search for a proposed signature of Lorentz-invariant spacetime granularity'', \textit{Class. Quantum Grav.} \textbf{28} 145011 (2011).

\bibitem{Petrov} A.Z. Petrov, ``The Classification of Spaces Defining Gravitational Fields'', \textit{Gen. Rel. Grav.} \textbf{32}, 1665 (2000).

\bibitem{Zee} A. Zee, \textit{Quantum Field Theory in a Nutshell}, (Princeton University Press, 2003), Chapter V3 page 267.

\bibitem{Ashtekar:2001}
A. Ashtekar, ``Quantum Geometry and Gravity: Recent Advances," [arXiv:gr-qc/0112038].

\bibitem{Jacobson1}
B.L. Hu and T. Jacobson, ``Directions in General Relativity," (Cambridge University Press, UK, 1993), page 364.


\end{thebibliography}
\end{document}